# Can female fertility management mobile apps be sustainable and contribute to female health care? Harnessing the power of patient generated data

## Analysis of the organizations active in this e-Health segment

**Maki Miyamoto, Rome Business School; and L-F Pau, Copenhagen Business School**

lpau@nypost.dk

**ABSTRACT**

**Is analyzed  the eHealth segment of female fertility planning mobile apps (in US called: perio d trackers) and its possible extensions to other female health care mobile services. The mark et potential is very large although age segmentation applies. These apps help women record and plan their menstruation cycles, their fertility periods, and ease with relevant personaliz ed advice all the uncomfort. As an illustration, the case of a European app service supplier is described in depth. The services of ten worldwide suppliers are compared in terms of functi onality, adoption, organization, financial and business aspects. The research question: "Can f emale fertility management mobile apps be sustainable and contribute to female health car e?" is researched by a combination of academic literature study, testing of 7 essential hypot heses, and a limited user driven experimental demand analysis. Quality and impact metrics f**

**rom a user point of view are proposed. The conclusion is a moderate yes to the research question, with several conditions. Further research and innovative ideas, as well as marketing and strategic directions are provided, incl. associations with male fertility apps.**

TABLE OF CONTENTS

**Document includes: 7 Figures and 6 Tables**

## 1 INTRODUCTION

In recent years, personal health technologies have emerged that allow patients to collect a wide range of health-related data outside the clinic. These patient-generated data (PGD) reflect patients' everyday behaviors including physical activity, mood, diet, sleep, and symptoms (Choe E. K. et al, 2018). However, major players and academics alike, have ignored the case where these patients or normal people are women!

As everyone knows, females are over a long segment of their lives, subject to fertility cycles which affect their lives. These cycles cause also periodic bleedings, pains, headaches etc., but they are subject to many external factors and personal lifestyle exposures (Collins S., Arulkumaran S., Hayes K., Jackson S., Impey L. (Eds), 2016). Recently some organizations and start-ups have attempted to allow individual females to record data about themselves so that they can plan in a responsible way activity as well as measures such as contraceptives.

With the widespread use of smartphones, which are personal and personalized, and their very large user base, several organizations and start-ups have over the past 10 years tried to enter the female specific e-Health segment (Eren H., Webster J.G.,2015) (Misra S., 2015) (GSMA Association, 2018).

However, as female fertility management is a novel eHealth application field, with strong elements of privacy and intimacy (Brands S.,2003) (Rajindra A., Richards D., Scott K.,2014) (Dehling T., Gao F., Schneider S. Sunyaev A.,2015) (Women's Forum,2018), it raises research questions on the choice of the delivery vector, of the targeted pains / illnesses, on the validation by e.g. gynecologists, of risks generated, and of course of cost & benefits.

The following Section 2. will illustrate this Introduction through a portrait of an app supplier organization in that segment.

## 2. CASE OF A FEMALE FERTILITY PLANNING APP SUPPLIER: BioWink GmbH

The German company BioWink (BioWink, 2018a) (BioWink, 2018d) is a prime source for researching the fertility mobile app segment because they have some years of notoriety and they have a European base. However, they refused providing any information on their organization, and refused any interview. Likewise, Glow, Luna Luna also declined, and My Cycle did not answer. Thus, this Section is based solely on external information, and serves only as a Case in the analysis of organizations providing such mobile apps.

BioWink GmbH, Berlin provides a Web app as well as a mobile application for use in iPhones and Android smartphones to track the fertility cycle. It offers a service around the mobile app named Clue, which enables users to track their monthly cycles by entering data about period, pain, mood, fluid, sexual activity, and personal notes.

The app is claimed by BioWink to have over 2.5 million users in 180 different countries (timing unknown), the very clear majority of whom make a free use of it. The startup has raised $10 million. Prior to a capital infusion in 2016 they claimed to have over 5 M users; the total investments should have reached $30 MUSD led by Nokia Growth Partners, along with existing investors Union Square Ventures, Mosaic Ventures, Brigitte Mohn and Christophe Maire, and additional new investors, Giving Wings and FJ Labs (BioWink, 2018a).

*"Clue has already a wide following amongst mostly young women (age group 15-45 years). For them, the most significant benefits are: a reduction in the disagreements due to menstruation, better diagnosis of other common illnesses, fertility planning, etc. … It is one of a small set of similar mobile apps for the benefit of young women's health problems (fertility, gynecologic pain, AIDS prevention, etc.).More specifically Clue provides a period and ovulation time reporting on a personalized basis (www.helloclue.com)."*

*"You can look at many female knowledges related to menstruation on the Clue website each in different categories, based on research, and cooperation with top research institutions and clinicians to explore topics with real-world impact — to better understand our bodies and physical mechanisms, explore our cultural and demographic diversity, and to break a harmful taboo. We work with carefully chosen researchers within academic institutions to answer specific research questions of a non-commercial nature. Research collaborating with Colombia University's Biomedical Informatics department, Stanford University's research scientist in the field of computational biology Department and computer scientist, University of Oxford and so on.*

*They are educating many different categories by knowledge from the perspective Anatomy, Bleeding Variation, Diet & Exercise, Emotions (PMS&PMDD), Endometriosis, Fertility & Menstrual Cycle, illness, Trans & Health, Vaginal & Cervical Fluids. Not only basic educational information but also focus on life stage for women such as Menopause, Pregnancy, Birth & Postpartum, Puberty. Especially could be interesting aspect Clue website is also regarding to Culture such as gender equality, tampons, Pads. Clue website gives significant information also long living life term for women according to sex life, Birth control, dating, SIT (sexually transmitted infections) which needs understanding and corporations with partner and partnership.*

*One of the most important things in this application is that women can obtain meaningful health information from websites with or without registering applications. And by further registering the application and entering information, information of the whole world is accumulated, privacy is protected based on legal regulation of personal information protection, while not performing any marketing or information trading etc.., every person It is not an exaggeration to say that the information is contributing to the research analysis of women's health is a long-term and innovative e-health strategy."* (Clue Privacy Policy (BioWink, 2018b)).

The founder and CEO of BioWink is Ida Tin, from Denmark, an adventure-writer credited with launching the term "femtech", and recipient of "European Female Web Entrepreneur Award" but with no background in medicine or finance or IT.

Regarding financials, at their start BioWink exhibited very high self-confidence as illustrated by the following quote, in which statements are not verifiable nor consistent with other information above from same supplier:

*"We have taken on venture capital from funds and people we trust, like Nokia Growth Partners, Union Square Ventures, Mosaic Ventures and many more. We chose to take investment—$30 million to date—from people who share our vision of a world where menstruation and reproductive health aren't taboo topics, but an accepted and essential part of life."* Press reports state BioWinks has about 50 employees, but this is hardly verifiable, esp. as even the company telephone number does not answer (from April 2018).

Regarding branding, the following is claimed (BioWink, 2018c):

*"Non-intrusive advertisements within the Clue app you have the option to remove with a paid subscription. An ad network like Facebook or Google might be used to display these ads. These networks track your browsing history to show you ads based on your interests.*

*There is an* online store *where you can buy Clue merchandise."*

## 3. FUNCTIONALITIES AND COMPETITIVE LANDSCAPE FOR FEMALE FERTILITY MANAGEMENT APPS

Obviously, there are other organizations and start-ups dealing with mobile apps for female fertility management than BioWink GmbH. Historically, almost all the basic knowledge resides with gynecologists and corresponding research centers in gynecology (US Veterans Administration (US) (US Dept Veterans Affairs, Undated), (National Health Planning of India National Health Planning of India,2017), Medstar Health Research Institute (USA) (Medstar Health Research Institute,2018), INSERM (Vaiman, 2018), etc..

However, it is the entrepreneurial spirit who has led to prototyping and even to the creation of start-ups found round the world. Table 1 below surveys such major organizations and their mobile apps, highlighting the application name, deployment languages, and some elements of functionality.

Significant work has gone into identifying from the user interface analysis of the major mobile apps the following classes of supplied app functionality:

① *Menstruation prediction control, checking*

② *Ovulation prediction checking*

③ *Basal body temperature*

④ *Making note of specific physical condition*

⑤ *Diet functionality such as weight control*

⑥ *Predictive Notification for next menstruation*

⑦ *Cycle extent education (information about sleepiness, irritation, etc.)*

⑧ *Health log (food, feeling, Sex drive status)*

⑨ *Symptoms (headache, digestion)*

⑩ *Sexual activity*

⑪ *Medication log*

⑫ *Fertility assistance*

⑬ *Information, Q&A, articles*

⑭ *Messaging and chat in app*

⑮ *Physical training program when pregnant*

⑯ *Emotional mood*

⑰ *Pill diaries*

⑱ *Vaginal condition (vaginal secretions)*

This functional analysis of supply is a contribution of this article, and is reused throughout, with the same functionality numbering.

**Table 1: Competing mobile apps for female fertility management**

| | ***Application (Country)*** | ***Developer*** | ***Deployment languages and functionalities*** |
|---|---|---|---|
| *1* | *Clue*<br>*(German)* | *BioWink GmbH*<br>*https://helloclue.com* | *Languages: English, Danish, French, German, other total 17*<br>Functionalities: ①②③④⑤⑥⑨⑩ ⑪⑯⑱ |
| *2* | *Flo*<br>*(US)* | *OWHEALTH, Inc.*<br>*https://flo.health/about-us/* | *Languages: English, Danish, French, German, other over 23*<br>Functionalities: ①②③④ |
| *3* | *Eve*<br>*(US)* | *Glow, Inc.*<br>*https://glowing.com* | *Languages:*<br>English, Portuguese, Chinese, Spanish<br>*Functionalities:* ①②④⑤⑥⑦⑧⑨ |
| *4* | *Glow*<br>*(US)* | *Glow, Inc.*<br>*https://glowing.com* | *Languages: only English*<br>*Functionalities:*<br>①②④⑥⑨⑩⑪⑫⑬⑭⑮⑰⑱ |
| *5* | *Period tracker Lite*<br>*(US)* | *GP International LLC*<br>*https://gpapps.com* | *Languages: English, Dutch, French, German, other 17*<br>Functionalities: ①⑥ |
| *6* | *My Cycles*<br>*(US)* | *Medhelp*<br>*https://www.staywell.com* | *Languages: English, Japanese*<br>*Functionalities:* ①②④⑤⑥⑨⑯⑰⑱ |
| *7* | *Cycle*<br>*(Sweden)* | *Perigee AB*<br>*http://perigee.se* | *Languages: English, French, German, other 12*<br>Functionalities: ①②③④⑥ |
| *8* | *Selene Calendar*<br>*(Japan)* | *Yahoo Japan Corporation (no app Web site)* | *Language: Japanese, Chinese, English, Korean*<br>*Functionalities:* ①②④⑤⑥ |
| *9* | *Luna Luna*<br>*(Japan)* | *MTI Ltd*<br>*http://www.mti.co.jp/eng/* | *Language: Japanese, English*<br>*Functionalities:* ①②③④⑤⑥ |
| *10* | *Me memo*<br>*(Japan)* | *Ayumi Co. Ltd. (no app Web site)* | *Language: Japanese, English*<br>*Functionalities:* ①②③④⑤ |

Reference (Kobanami, Yoneda, Ali, 2018). mentions briefly some additional apps for period tracking, such as: Period Calendar by Abishking Ltd (predecessor of Period tracker), and My Calendar by Emily Powell, which have a very limited following and only on iPhone.

This allows us to determine in Table 2, the most frequently *supplied* functionalities in those 10 currently available dominant mobile apps from Table 1.

A later analysis in Section 8.5 and Appendix 1 will show what is the functional *demand* counterpart from users.

**Table 2**: Distribution of functionality supply in Table 1 mobile apps for female fertility management

| ***Functionality in supply*** | ***Description*** | ***Percentage of suppliers offering that functionality (in %)*** |
| --- | --- | --- |
| ***1*** | *Menstruation prediction control, checking* | ***100%*** |
| ***2*** | *Ovulation prediction checking* | ***90%*** |
| ***3*** | *Basal body temperature* | ***50%*** |
| ***4*** | *Making Note of physical condition* | ***90%*** |
| ***5*** | *Diet functionality such as weight control* | ***60%*** |
| ***6*** | *Predictive notification for next menstruation* | ***80%*** |
| ***7*** | *Cycle extent education (information about sleepiness, irritation, etc..)* | ***10%*** |
| ***8*** | *Health log (food, feeling, sex drive status)* | ***10%*** |
| ***9*** | *Symptoms (headache, digestion)* | ***40%*** |
| ***10*** | *Sexual activity* | ***20%*** |
| ***11*** | *Medication log* | ***20%*** |
| ***12*** | *Fertility assistance* | ***10%*** |
| ***13*** | *Information, Q&A, articles* | ***10 %*** |
| ***14*** | *Messaging and chat in app* | ***10%*** |
| ***15*** | *Physical training program when pregnant* | ***10%*** |
| ***16*** | *Emotional mood* | ***20%*** |
| ***17*** | *Pill diaries* | ***20%*** |
| ***18*** | *Vaginal condition (vaginal secretions)* | ***20%*** |

We can conclude that all major suppliers have 2-3 functionalities, others have more, but are limiting themselves for fear of mobile app complexity and user data entry workload.
It is important to see, in view of user requirements, that interaction with other users (functionality 14) is limited to 10 % of apps on the market (essentially only: Glow).

## 4. DESCRIPTION OF EACH FEMALE FERTILITY MANAGEMENT MOBILE APP AND THEIR SUPPLIERS

The descriptions below are as citations the mobile app presentations by the suppliers in Table 1 (not the descriptions of corresponding Web based apps). As a common characteristic, they are very easy to read, refer to advanced IT tools without saying which is used for comparison, and never allude to service supplier liability and prediction errors. These are thus common characteristics of the *corporate and marketing focused communication strategies* of these suppliers, and the texts are provided as evidence of this approach. The claims to fertility accuracy prediction are often unsubstantiated and could hardly be established scientifically owing to many specific situations.

Some apps have an alert function, and/or an advisory function, and even community forums (see Table 1).

### 4.1. Clue

*"Clue has been ranked as the top period and ovulation tracking app by the Obstetrics & Gynecology journal, (Evaluation of Smartphone Menstrual Cycle Tracking Applications Using an Adapted Applications Scoring System, June 2016 which is a publication of the American College of Obstetricians and Gynecologists)* (Note of authors: full reference is (Moglia, M. L., Nguyen H., V., Chyjek K., Chen K., Castaño P.,2016))*: Clue uses science to help its users to identify unique patterns in their menstrual cycle. With the app's period tracker, multiple mood trackers, health logs, and exercise trackers, it is claimed that user's health and menstrual cycle will no longer be a mystery. The developers promise to be inclusive of all ages and never use butterflies, flowers, euphemisms, or pink. The app's unique algorithm learns from the data that user add, which means that the more you use Clue, the "smarter" it will become".* This tends to indicate that a simple neural networking induction algorithm is used.

### 4.2. Flo

*"If you are wondering when you last had a period or would like to know when your next one is due, you can easily find out using Flo. Flo uses machine learning to predict menstruation and ovulation. Using the app's bold and simple calendar, you will be able to log how you are feeling, your symptoms, sex drive, and menstruation flow. The app can also be used to track sleep, water consumption, and physical activity. The Insights dashboard helps you to learn more about your body and cycle, and it also provides personalized health insights each day."*

### 4.3. Eve

*"Eve is a period tracker that predicts upcoming periods and your chances of pregnancy. Discover trends in your cycle by logging your moods and symptoms and view your health data in eye-catching charts. With Eve, you can review past periods, forecast future periods and ovulation, and visualize your cycle history in a new way with Eve's interactive staircase. The app provides access to a community where you can discuss periods, sex, and health. If you are concerned about anything at all, ask the community. No subject is off limits, and you can learn from those who have had similar experiences."*

### 4.4. Glow

*"Glow can track your period and record your symptoms, mood, sexual activity, and medications. Glow's data-driven menstrual and ovulation calculator helps women to take control of their reproductive health. The app can forecast periods and ovulation and its predictions become smarter over time. Not only can the app help women who are avoiding or attempting pregnancy, but it also helps those who are undergoing fertility treatments such as intrauterine insemination and in vitro fertilization. You can make charts of your menstrual and fertility data, set medication, birth control, and ovulation reminders, as well as log more than 40 different health signals. Glow*

*also offers a subscription to unlock comparative insights, premium articles, private messaging, and premium support."*

Glow has launched a "Glow Fertility Program", combining knowledge from clinics and hospitals, with data collection from female users via the app. It involves about 40 clinics in 17 American States, in addition to recommendation doctors (Glowing.com, 2018).

### 4.5. Period tracker Lite

*"The Period Tracker Lite app makes logging menstrual cycles quick and easy. Press a button at the start of each period and Period Tracker Lite will record your data and use the average of 3 months' worth of data to calculate your next period. Take notes each day about your symptoms, such as flow, cramps, bloating, backache, headache, and tender breasts. Your weight, temperature, and a choice of more than 30 moods can also be selected. Period dates, fertility days, and ovulation are all shown in a simple month-view calendar. The app provides comprehensive charts that illustrate weight changes, temperature, period length, cycle length, and symptoms."*

### 4.6. My Cycles

*"Regardless of whether you are trying to conceive or become an expert on when your next period is due, My Cycles can help. My Cycles tracks periods in a handy calendar and predicts future ones. Wherever you are, you can track your period, symptoms, mood, and medications with My Cycles in an instant. Record your periods, view them at a glance with the easy-to-read calendar, and plan periods, fertile days, and ovulation for the next 12 months. If you are trying to get pregnant, the app lets you know when your chances of conceiving are higher with helpful reminders. Likewise, it tells you when to use extra protection if you are not trying to conceive."*

### 4.7. Cycles

*"The Cycles app is a simple period and fertility tracker that requires little input. Irrespective of having regular or irregular periods, with Cycles, all you need to do is turn the dial on the first day of your period, and that's it — the app automatically adjusts your cycle length. Cycles is designed so that you can invite your partner to the app to keep up with your cycle. This feature is useful so that your partner can provide emotional support, plan trips and romantic evenings, and know when your most fertile days are if you are trying to conceive. The app uses scientifically backed fertility tracking that predicts fertility with an accuracy of up to 95 percent. Optional password protection keeps your information secure from prying eyes."*

### 4.8. Selene Calendar

*"Selene Calendar is simple, easy to use and girly design menstruation tracking application. It could be recommended to younger generation women who is starting menstruation. You can write your comment of your body condition on comment page. Not only you can look and track in the calendar view for predicting the period data, ovulation data, possible conception date as icons clearly but also make various arrangements and holiday plans easily. For examples you can record the condition of your body and event, etc., with cute stamps. Additionally, you can also record weight for your diet and period and diet are closely linked to each other in daily cycle. "*

### 4.9. Luna Luna

*“This application is designed for all generation women by changing switch easily from contraception mode to pregnancy mode. You can recognize your status of notification of PMS moods, active mood after menstruation and so on by related to menstruation cycle. You can also look many different types of columns suits to your body condition which informed PMS or menopause problems and anti-aging beauty information. There are some alert functionalities for Best Diet dates through menstruation cycle.” Luna Luna medico” service which connect women who has gynecological problems and Doctors is useful communication functionality in this application.”*

This app is the dominant one in Japan, and is on trial at 4 hospitals, aiming for fielding at 1000 medical facilities by 2020 (Ooshita, 2017).

### 4.10. Me memo

*“You can use this application not only as tracking application, but also “personal recording like diary” by using many different functionalities such as recording: body temperature, weight, body fat, sleeping, meal, condition of skin, free memo, physiology basic temperature note and so on. If you start fill in your menstruation information, application would predict appropriate diet date, ovulation data, menstruation starting data. You can look your body temperature and body weight, sleeping hours by bar graph. Thanks to many different types of column like beauty, diet, meal, skin care information, many women could enjoy reading while using this application.”*

## 5. ACADEMIC LITERATURE SURVEY

There are only very few academic or similar articles on female fertility management apps. A few have been selected below for analysis. There are however very many books and articles on mobile business, which we will not survey here as they bear little specific relevance to this research (e.g. (Information resources Management Association USA, 2018)).

### 5.1. Eric Wicklund, mHealth Study Ties App to Improved Outcomes for Pregnant Women (Wicklund,2017)

A Medicaid-based mHealth study in Wyoming has found that pregnant women using a customized app are far more likely to consult with doctors during maternity – and they're more likely to deliver healthy babies.

The study, conducted by the Wyoming Department of Public Health in 2014-15 using the WYhealth Due Date Plus app from Wildflower Health, found that app users were 76 % more likely to schedule prenatal visits at least six months before delivery than those not using an app.

More importantly, pregnant women using the app were only 25 percent as likely to deliver a low-birth-weight baby as those women who didn't use an app.

*"Wyoming Medicaid saw fewer low-birth weight children in women who used the app than those in Medicaid not using the app, and we feel the app produced a 3:1 ROI (Return on investment) based on cost avoidance,"* James Bush, MD, FACP, Wyoming Medicaid's medical director and the study's co-author, said in a press release. *"Cost savings and improved birth outcomes is a winning combination."*

The study focused on 85 pregnant women on Medicaid who used the app and another 5158 pregnant non-app users with delivery outcome records in Wyoming Medicaid. More than 1730 people reportedly downloaded the app during the study period, with registered Medicaid members using it an average of 6.4 times a month.

According to the study, in addition to short-term costs for delivery and perinatal care, negative pregnancy outcomes such as preterm birth and low birth weight carry important long-term consequences.

*"Prematurity is a major driver of pediatric morbidity and disability, and it is associated with nearly 50 percent of all childhood neurodevelopmental disorders,"* the study's authors reported. *"In a 2006 report, the Institute of Medicine (IOM) estimated an annual societal cost of $26.2 billion ($31.8 billion in 2015 dollars) associated with preterm birth in the United States, of which 2 percent stemmed from early intervention services, 4 percent from special education services, and 22 percent from lost household and labor market productivity."*

*"Research aimed at improving U.S. birth outcomes has found that one of the most important determinants of birth outcomes is access to prenatal care,"* the report notes. *"Studies have found that even after adjusting for other differences like socioeconomic status and maternal age, infants born to mothers who receive no prenatal care weigh considerably less, on average, than those whose mothers receive prenatal care. In contrast, early prenatal care in the first trimester (the first 90 days) has been shown to have a significant positive effect on birth outcomes; for instance,* a *sample of white teenagers showed a 27 percent reduction in low-birth weight births when provided with early prenatal care."*

The app connects pregnant women to clinical information on common pregnancy symptoms and key health milestones. It also links users to Medicaid benefits information and community-based health and wellness resources such as the Women, Infants and Children and Wyoming Quit Tobacco programs.

*"This study demonstrates that mobile maternity applications can positively influence people's decisions to attend early prenatal appointments and improve their connections to healthcare resources,"* Dilek Barlow, MA, director of client services at San Francisco-based Wildflower Health and a co-author of the study, added in the press release. *"Moreover, it shows that Medicaid populations can be highly engaged with digital tools that have the potential to greatly improve outreach and education in this population."*

### 5.2. **Kevin Anderson, Oksana Burford, Lynne Emmerton, Mobile Health Apps to Facilitate Self-Care: A Qualitative Study of User Experiences** (Anderson, Burford, Emmerton, 2016).

Twenty-two consumers (15 female, seven male) from Australia participated, 13 of whom were aged 26–35 years. Eighteen participants reported on apps used on iPhones. Apps were used to monitor diabetes, asthma, depression, celiac disease, blood pressure, chronic migraine, pain management, menstrual cycle irregularity, and fitness. Most were used approximately weekly for several minutes per session, and prior to meeting initial milestones, with significantly decreased usage thereafter. Deductive and inductive thematic analysis reduced the data to four dominant themes: engagement in use of the app; technical functionality of the app; ease of use and design features; and management of consumers' data.

The reference deals with self-care applications and consumer experience metrics. Due to differences between chronic conditions, there is no agreed definition of self-care. One commonality is that self-care requires *self-monitoring* for a consumer to pursue daily decisions to maintain functionality. Self-monitoring can be conducted by consumers on various levels; examples are self-awareness of symptoms (e.g. shortness of breath), manual blood pressure readings, and self-maintained electronic databases of blood glucose measurements in diabetes management. For consumers with reasonable health literacy, self-monitoring offers greater autonomy, aiming to reduce pressure on health resources.

However, limited regulation in the technology marketplace enables insufficiently tested self-monitoring devices such as mobile apps to be launched, with potential for health consumers to ill-advisedly change their self-care regimens. There are many instances of 'buggy' health apps.

Studies into self-care using mobile apps have predominantly involved custom-designed apps. In these studies, self-efficacy was the only measurement of consumer experience, while participants' engagement with the app was determined via self-reporting. Engagement does not necessarily mean long-term commitment to using the app; therefore, combining such data with usage statistics, such as login time and frequency and accessed features would add value to these studies.

Notable deficiencies collectively demonstrated in these studies are their relatively short follow-up periods and lack of detailed consumer experience findings. Additionally, self-management programs have measured select outcomes, rather than a more holistic spectrum of outcomes relevant to conditions such as diabetes, osteoarthritis and hypertension.

The Mobile Application Rating Scale (MARS) is an Australian development from 2015 to assess the quality of health apps, and is based on four quality scales: engagement, functionality, aesthetics and information quality. Research applying the MARS in studies involving health apps is emerging, with MARS noted in an Australian wellbeing evaluation protocol and mentioned in an Irish mental health app feasibility study without being used in the study itself.

Relevant to this research, the paper refers to limited experimentation and survey of menstrual cycle monitoring, although the female respondents were few (Table 3).

**Table 3**: From Reference [H]: consideration given by 4 surveyed persons to menstrual cycle monitoring

| Type of App | Used by Android Participant Number | Used by iOS Participant Number | Number of Participants |
|---|---|---|---|
| Blood pressure monitoring app (1 type) | | P6 | 1 |
| Diabetes monitoring app (2 types) | | P2, P17, P20 | 3 |
| Migraine management app (2 types) | | P5, P8 | 2 |
| Menstrual cycle monitoring (4 types) ⇦ | P1, P22 | P6, P4 | 4 |
| Anxiety management app (1 type) | | P13 | 1 |
| Calorie management and weight loss monitoring app (5 types) | P1 | P2, P3, P16, P20 | 5 |
| Celiac disease management app (1 type) | P11 | | 1 |
| Sleep monitoring app (4 types) | P14 | P6, P13, P21, | 4 |
| Pain management app (2 types) | | P8 | 1 |
| Cycling app (2 types) | | P12 | 1 |
| Fitness App (22 types) | P8, P9, P11, P14, P18, P22 | P2, P3, P7, P9, P10, P15, P16, P17, P19, P20, P21 | 17 |
| Other (saliva analysis kit) | | P16 | 1 |

A user of a menstrual cycle tracker (label P4) was not comfortable with the prospect of her data accessed by third parties, while another was less concerned:

*"I don't think about [data security], to be honest. This is going to sound terrible—maybe I'm just naive … I don't know, it doesn't really concern me. Probably, it should."*

Perceived benefits from usage of health apps included desire to customize app features to suit individual needs. Participants also expressed greater control of their condition, in this case, menstrual problems:

*"I decided just to search and find out whether there was an appropriate app just to make life a little bit easier… my specialist had told me to keep track of any symptoms and the length of my cycle, so I just found [the menstrual cycle tracker] myself online* and *found that to be an easy tool to use."*

*"Maybe if I could leave the features I don't use [in this menstrual cycle tracker] behind, since I'm not trying to get pregnant, so just get rid of these fertile days."*

The present data cannot conclusively support the correlation between "willingness to pay" and "user experience", although the correlation has been reported elsewhere in a study of mobile apps (Hebly, 2012).

Participants tended to reduce usage of their app when they reached their goals and no new self-management techniques were offered. For app engagement to be sustained after reaching a goal or for usage to become habitual, regular intervals of engagement were recommended.

### 5.3. **S. Storey, Y.Y. Yang, C-L Dennis," A mobile health app based postnatal educational program (home-but not alone): descriptive qualitative study** (Storey, Yang, Dennis, 2018) and (Eysenbach, 2018)

This research on an mHealth app-based postnatal educational program is related to fertility planning app´s knowledge dissemination (functionality no 7 in Clue, Eve and Glow) for women from adolescence to menopause. In that educational program, audio recordings and video are provided, accentuating the accessibility of information to the user as well as healthcare professionals. Also, the postnatal educational program described in the article, includes face-to-face interaction. The fertility planning apps used did not include these technical and expert interaction channels.

It is noted that psychological anxiety to the female user is reduced by face-to-face dialogue; however, the apps simultaneously represent a time saving measure as they reduce the burden on hospital or medical staff.

## 6. ANALYSIS of the FEMALE FERTILITY PLANNING e-Health SEGMENT

This Section extends the academic literature review of the previous Section 5, by analyzing industry, economics, public health, government and financial data not reported by academics.

### 6.1. Benefits and impact: Extent of eHealth services provided and drivers

First, we can confirm that mobile apps for female fertility period management, do contribute to, and are part of eHealth. Taking as a basis the WHO definition (Figure 1), they do not however satisfy all criteria, but only some of them:

a-ICT serves women's health in the medical disciplines of gynecology, family planning, nutrition, and oncology;

b-ICT rests upon an exchange of user generated information, assumed to be complete, truthful and linked to the individual user, and the mobile app data collection platform;

c-ICT does not always (see Section 4) provide distance health care services, nor consultant advice by health care professionals;

d-Female fertility management ICT does not show how continuing education of health care providers is carried out, as the mobile app providers do not make explicit if and how the data collected is passed on to these professionals, and under which privacy rules.

**Figure 1**: WHO eHealth definition (http://www.who.int/topics/ehealth/en/)

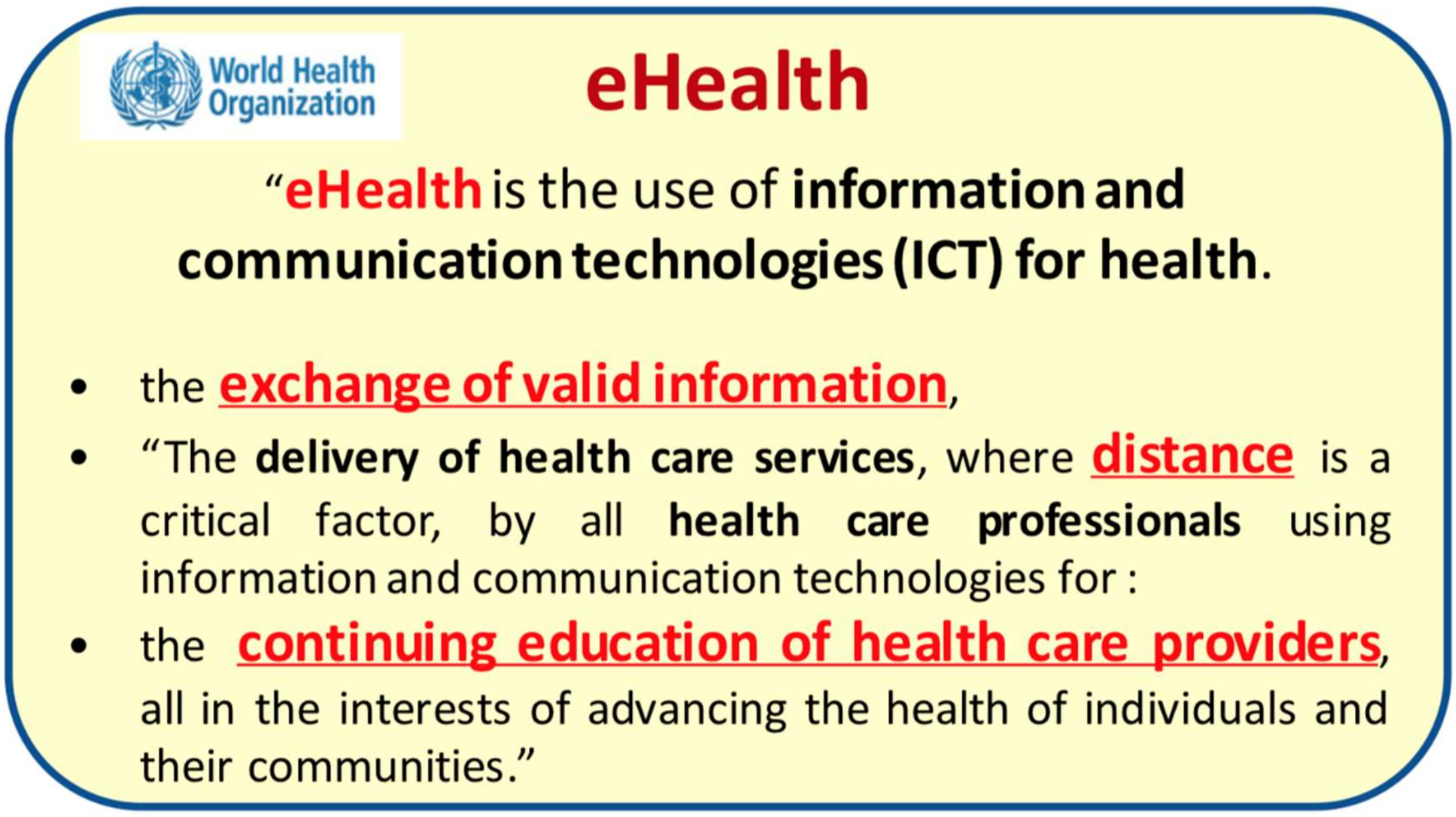

A general analysis framework for female fertility management is provided in Figure 2. It relies essentially on gender research and women's empowerment, and not on IT or medical research. This point of view will be adopted below.

**<u>Figure 2</u>**: Health and MNCH state of the evidence empowerment analysis (Deshmukh, Mechael, 2013) (Philbrick,2013).

Figure 1.
mHealth and MNCH: Addressing Gender and Women's Empowerment—A Framework for Analysis

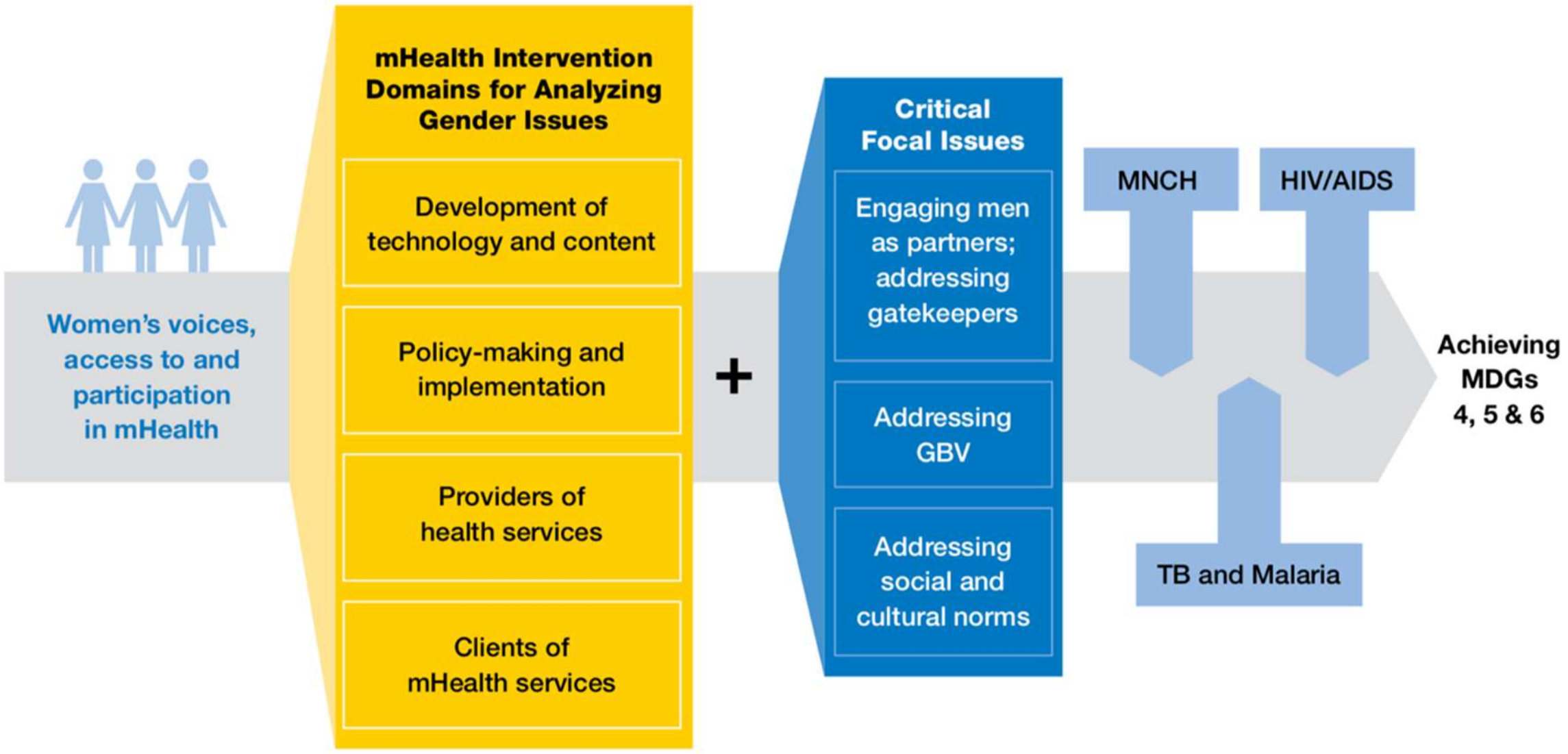

This general framework essentially drives women's demand for m-Health services, such as female fertility management, from key critical gender issues, including social roles, and roles in the labor workplace.

So, to further elaborate upon point a- above, while information is overflowing in the information society, it is necessary for women to focus on women's physiological cycle and to pick up and provide information, to allow women to be active in society and live busily. This demand drive is documented in Figure 3, which gives the female workforce distribution in the European Union in 2011. Figure 4 illustrates the male/female and family status breakdown in employment.

**Figure 3**: Percentage of women in workforce among all women aged 20-64 years in the European Union in 2011 (Wikipedia, 2018).

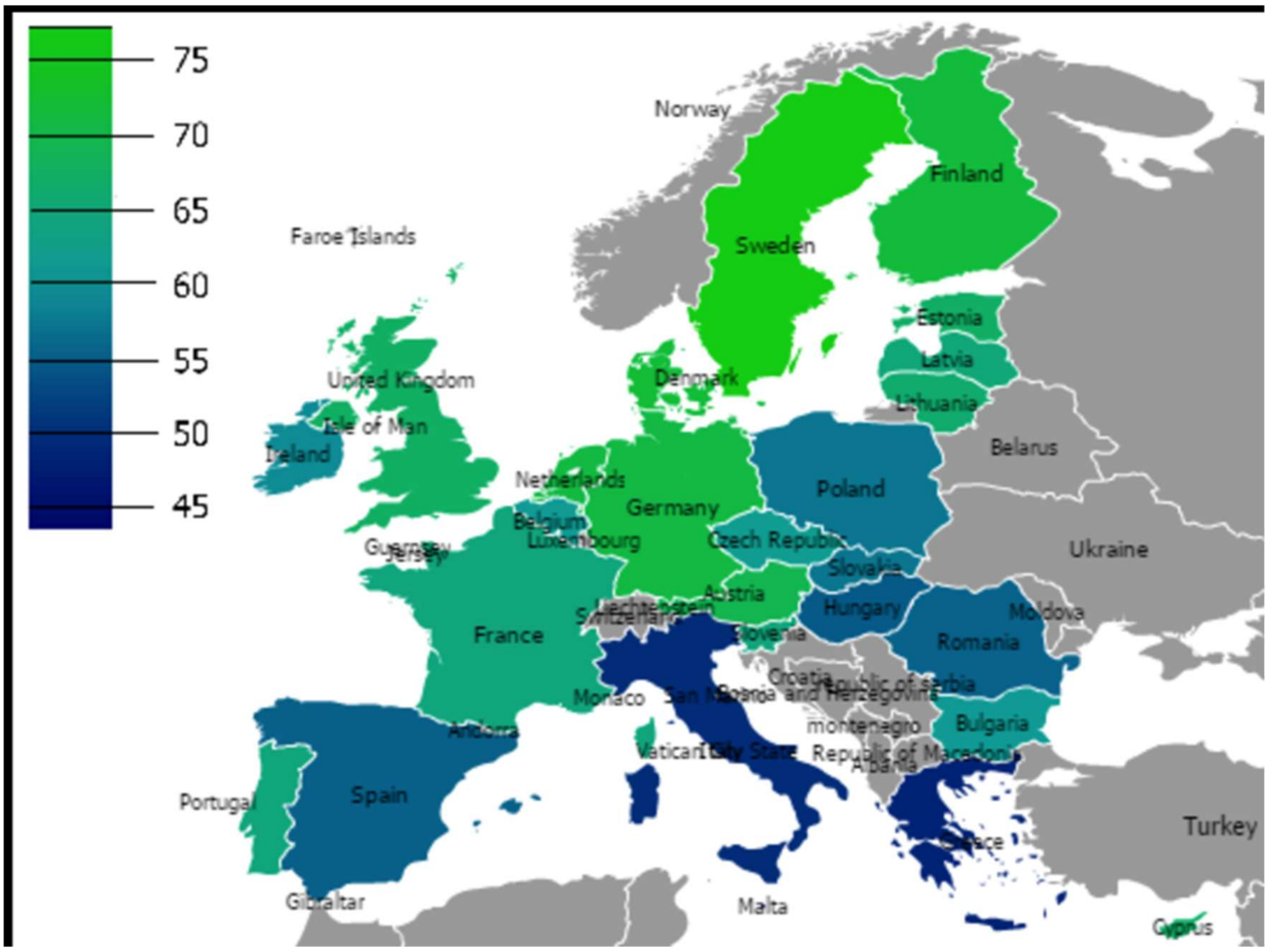

Some mobile app suppliers, such as Clue, claim (although such claims could not be verified as mobile app developer does not provide expert identities), that women can acquire knowledge with confidence and learn from the functionalities in Table 2. Considering social circumstances, female fertility management apps contribute to maintain a balance between both family and work obligations.

From the business point of view, the female gender accounts for almost half of the population, and this yields a very high addressable market share. If women's health improvement is made feasible with mobile applications, it is not an exaggeration to say that it will lead to reductions in medical costs in each country, reduction in working hours of medical staff, and social economic improvement.

**<u>Figure 4</u>**: International Labor Organization (ILO): Male/female and family status breakdown in employment in 2015 (International Labor Organization, 2016).

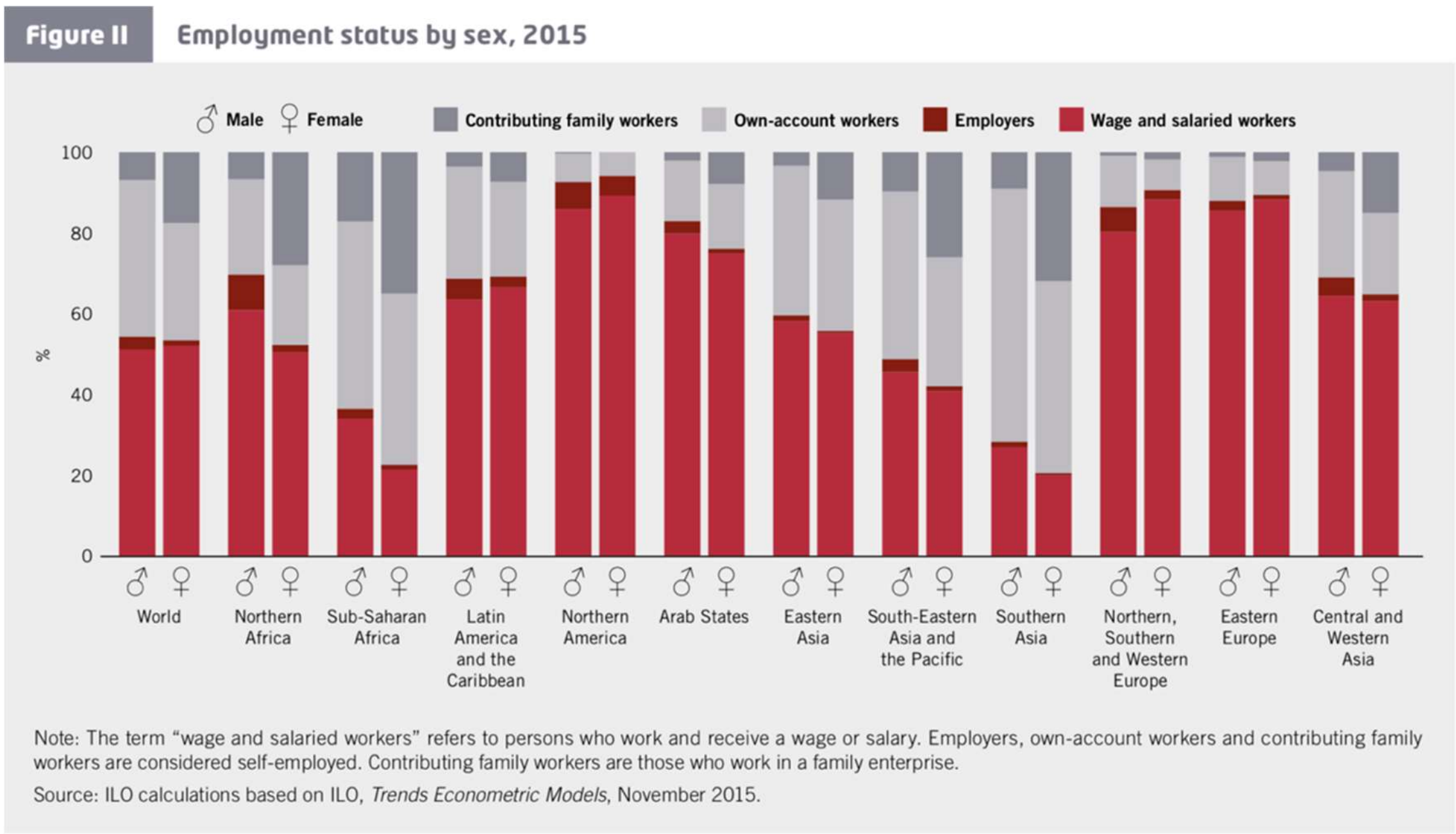

Women were largely limited to low-paid and poor status occupations for most of the 19th and 20th centuries or earned less pay than men for doing the same work. However, through the 20th century, public perceptions of paid work shifted as the workforce increasingly moved to office jobs that do not require heavy labor, and women increasingly acquired the higher education that led to better-compensated, longer-term careers rather than lower-skilled, shorter-term jobs.

The increasing rates of women contributing in the work force has led to a more equal disbursement hour worked across the regions of the world. However, in Western European countries the nature of women's employment participation remains markedly different from that of men.

Although access to paying occupations (the "workforce") has been and remains unequal in many occupations and places around the world, scholars sometimes distinguish between "work" and "paying work", including in their analysis a broader spectrum of labor such as uncompensated household work childcare, eldercare, and family subsistence farming.

The female fertility mobile apps are also transmitting information to women outside the menstrual period, such as women before and after menopause; in addition to that, they contribute to enhanced education and knowledge of young pre-menstrual young women or to related organizations.

### 6.2. Female fertility mobile app limiting factors and privacy

That female fertility management eHealth app segment suffers from the same general limitations as found in other eHealth services, as illustrated in Figure 5:

**Figure 5**: eHealth service limitations (from Innovation and implementation eHealth in the WHO European region in 2016) and barriers to implement MHealth programs (WHO, 2016).

Fig. 16. Barriers to implementing telehealth programmes

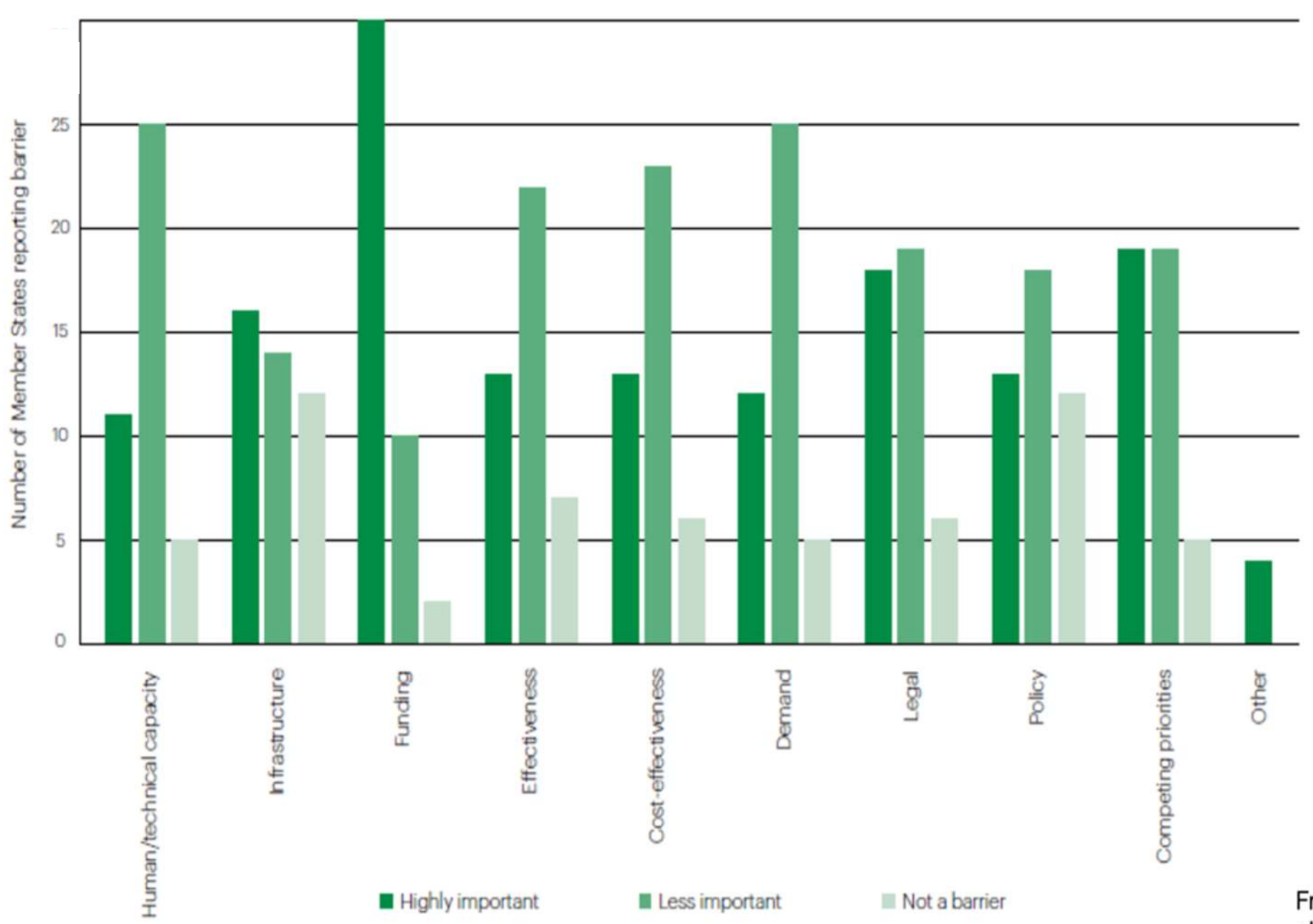

From Innovation to Implementation eHealth in the WHO European Region 2016

Barriers to implementing mHealth programmes in 2015

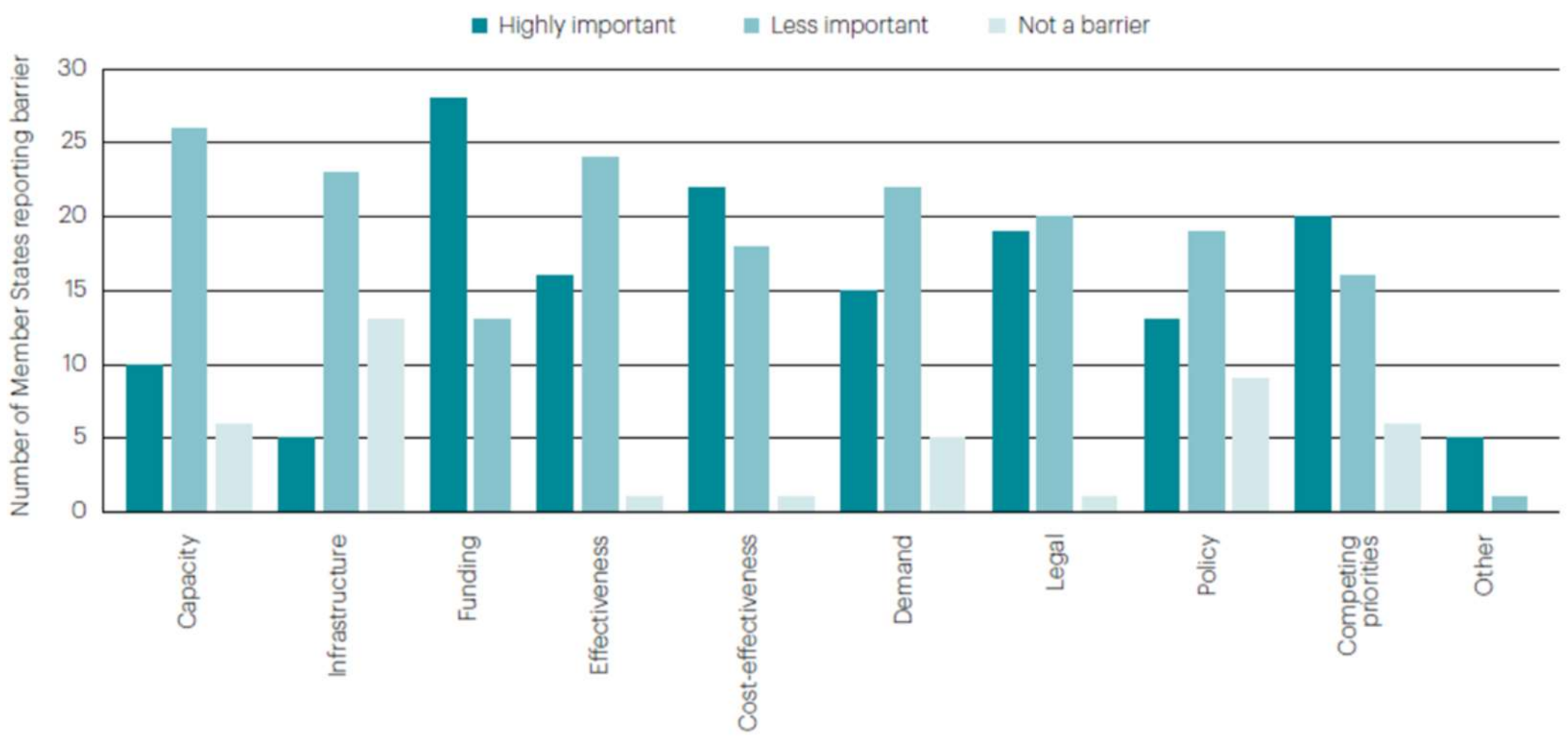

From Innovation to Implementation eHealth in the WHO European Region 2016

A long -term strategy is necessary for female fertility mobile apps deployment, to improve social trust and to respond to the needs of female users, before expanding the functionalities to other female health issues. Particularly problematic is that some suppliers such as those of Table 1 mediate Social Networking Services, thus there is a significant risk for privacy and personal information leakage due to an increase in application usage. In addition, general purpose social networking apps like Facebook, are subject to even wider leakage from their discussion groups. The problem will be discussed in detail below, discussing first a Case.

Case: Clue developers are aware of privacy and security risks, and try to quiet down user concerns by text such as (BioWink, 2018b):

*"At Clue, technology provides a groundbreaking opportunity to empower people to take control of their health. We fully acknowledge the great responsibility that comes with safeguarding sensitive data, such as information about your menstrual cycle. We are committed to achieving the highest standards of privacy and security, as well as to being transparent about how we process data. Clue is made in Berlin, Germany, and the way we handle data meets German legal requirements. Registering your health information for consumer users is the same as having the possibility of being exposed to the risk of leakage of personal information. Companies must inevitably accumulate and utilize information under strict management and responsibility against it. However, for consumers, utilizing this application function that can find more value while improving the reliability of personal information which is important not only for individual person's health management, but also for all their accumulated value, as companies argue it data is utilized for research analysis, that is, information of one individual can contribute to women's health information on a global scale."*

This case shows that the European regulations on General Data Protection (European Commission, 2016) is probably respected by some suppliers, but that they try to convince users about the benefits of pooling and diffusing data provided by users. Suppliers enable privacy compliance at a very minimum level as otherwise their business models would be in jeopardy, and they are anyway directly exposed to forthcoming European regulations on data aggregation and exploitation, which rely on individual and controlled end user consent to let others use, or not use, their data; see e.g. Clue Privacy Policy (BioWink, 2018b). In Japan, the user may authorize explicitly a hospital doctor to access the private data collected via the Luna Luna app and link them to the electronic medical record; this is a far stricter privacy policy.

While contributing to women's health care, there are specific problems and barriers for each female fertility applications, in addition to the general limitations discussed at the beginning of 6.2., even in the process of improving women's health. These specific additional problems and barriers include the following observations (see also Appendix 1 experimental research):

-women users think it is troublesome to enter the data in the app, especially as there is a time discipline involved (some days are more important than others during a period);

-it is visually difficult to see how one's own data fit into general curves or prescriptions: "where is my case?"

-the general awareness of female fertility apps is still low;

-there is a big competition from other applications, social media, discussion groups, concerned with female health in general.

## 6.3. Liabilities and Risks

The e-Health segment of female fertility planning apps is not risk free for obvious reasons. A female user could in principle sue the mobile app provider for providing erroneous period prediction for dates, or for prescribing incomplete measures to be taken based on the advice provided by the app, or for lack of notifications altogether (as reported in Appendix 1). Strangely enough, none of the apps suppliers from Table 1 include any legal liability provisions to these effects.

## 6.4. Organization and processes at suppliers of eHealth services for female fertility planning

### 6.4.1. Organization

All the suppliers of female fertility planning tools of Table 1 are very small start-ups, around a core of founders, and sometimes seed capital investors (including the founders). An extensive information search has not produced any sources describing aspects to the organizations of these suppliers, other than, when made public, the initial founder was a woman (this is the Case of Clue with lead founder Ida Tin, see Section 2.). As BioWink AGH refused any interview or visit, no primary data could be collected.

Thus, the assumed organization must be deducted from that of other eHealth start-ups which have in open fora explained their history; the case used is here: Doctossimo (Unnamed, 2018b). We also assume a size of employees derived from the business analysis in Section 8.6, showing that most female fertility planning mobile app suppliers have very limited staff if any.

The assumed core organization is:

-a founder or key partner specialized in gynecology, or family planning needs, and interacting with users, animating the user community Web page, and obtaining needed certifications;

-a founder or key partner in charge of operations, sourcing of mobile app development platforms, and development of the mobile apps (incl. their internationalization);

-an administrative assistant, in charge of contracting with mobile app host operator (in the cloud or not), revenue collection, selecting advertising media, managing contracts with social media like Facebook, etc.

This core organization is supplemented and supported by work subcontracted to:

-freelance software developers;

-a mobile app hosting operator (or cloud content management operator);

-marketing companies or specialized free-lance specialists.

In addition, the start-up usually relies on the exposure to financial media and advertising contacts which the seed capital fund and founders may provide. This is concretized by investors days and participation in selected meetings targeting users, investors, as well as the relevant medical professions (gynecology, family planning).

### 6.4.2. Processes

As in the previous subparagraph, no primary or secondary data could be found, so processes at suppliers of female fertility applications is derived by assumptions from cases of mobile app developers.

We identify the following main processes, many linked to the execution of the initial business plan, and they are consistent with what is reported by Doctossimo (Unnamed, 2018b):

1. From idea generation to eliciting app functionality (see Sections 2 and 4);

2. Software development and hosting of apps;

3. User metrics analysis and updates (using feedback, social media interaction, and possible data mining) (see Table 4);

4. Business revenue development by suitable usage plans, resale of ads, partnerships and eventual exploitation of anonymized user data (subject to privacy terms and user consent);

5. Cash flow analysis and financing (different seed and other funds, crowdfunding);

6. Marketing and awareness build-up, in association with investors;

7. Talent recruitment (when economically sustainable);

8. Quality control and security audits;

9. Strategic alliances;

10. Revisions of business plan.

### 6.5. Financial information regarding female fertility planning app suppliers

Obviously, there is almost no public information on the financials of female fertility management mobile app suppliers. Anyway, for entrepreneurial founders of such suppliers, and their investors, valuations rely on merger and acquisition transactions for similar products/services. Such other transactions allow to estimate market ratios and estimate even business levels based on independently collected user number information.

According to French business magazine Challenges (Unnamed,2018b), on 12 July 2018, the French no 1 TV and media company TF1, itself owned by the Bouygues conglomerate, purchased Doctissimo (eHealth advisory information service), and, in a parallel deal, “Mon Docteur” eHealth advisory site was purchased by Doctolib (eHealth doctor bookings) (average 1 M consultations/month). Both companies were sold by Lagardère, Europe’s largest printed media company.

Doctissimo is deployed in F, IT, SP and had in 2017 an audience of 12 Million registered users with 40 Million consultations /month. It was created in year 2000 and was owned by Lagardère since 2008 until this deal. It is focussed on health, nutrition and family matters and all health applications have mobile app instances. Together with its previous purchase of Doctolib, Bouygues / TF1 through Doctissimo becomes by far no 1 in Europe for eHealth bookings.
The analysis of the corporate announcement by Bouygues reveals the strategic vision it has is to capture dedicated media channels for women. Not only did the Bouygues majority owned subsidiary TF1 acquire the eHealth providers Doctolib and Doctissimo as discussed, but in parallel TF1 directly acquired 78 % of the established large-audience aufeminin.com Web site in April 2018 at 39,47 Euro/share from the Axel Springer German media company (Unnamed, 2018b); aufeminin.com is catering to a broad range of topics for women. The director of innovation of TF1 was named CEO of aufeminin.com.
Thus, one can formulate the hypothesis that one day Bouygues will acquire a provider of female fertility planning applications, especially if this supplier has extended its uses to other female health issues.

The combined sales value of the two companies Doctissimo and "Mon Docteur" is reliably reported to be 60 M Euros (Unnamed,2018b). This last information opens for a financial analysis relevant for female fertility mobile apps valuation and business estimates.

Assuming from Google Analytics for "Mon Docteur" about 1 Million registered paying users, and reusing the audience metrics for Doctissimo, the reported deal value gives a going rate of 60 MEuros / (12+1 M registered paying users) = *roughly 5 Euros valuation by paying eHealth app registered user.*

The 10 MEuros investment in BioWink (see Section 2) at formation is in line with the above merger and acquisition valuation rates. Using the analysis in Section 8.5., and (Roche J.,2018), with for Clue a verified paying subscriber base in 2018 of 46 800 users, 10 years after BioWink 's creation, this gives for the venture funds who invested in BioWink a valuation of 10 MEuros / 46800 subscribers = *213 Euros of seed investment value per registered paying customer in a stable growth phase*, which is very high and reflects over expectations driven by the corporate communications (Section 4).

To analyse *business operations* from an eHealth female fertility service, we define in this article a *session* as the time span over which a given registered eHealth app user, eventually inputs data to this app and browses general information as well as processed personal information. The precious peculiarity of female fertility planning is that it is has a recurrent business periodicity, as opposed to many eHealth apps which serve only just before or during an illness.

As TF1, Doctolib and Docissimo don't say how long a registered paying user stays, one cannot estimate business income from a session in relation to usage, but it is surely low. For all usage during a one-month session only, then the average business revenue per one-month user session is 60 MEuros / (40+1 M consultations in a month) = *1,5 Euros average business revenue per one-* *month user session*. It is probably 6 time less if a female fertility app user selects a 6-month session prior to a pregnancy, that is an average business revenue per one-month user session revenue of 1,5 Euros /6 months = roughly 0,25 Euros.

These business operations data are in fact confirmed independently by the experimental research in Appendix 1, although these average business revenues per one-month user session do not eventually include side-revenues to the suppliers from advertising, sales of data, courses, etc.

### 6.6 Intellectual property

As currently the female fertility planning apps are only informative, and rely on common gynaecology knowledge, there is no intellectual property protection for suppliers other than the software and user interface copyrights, besides possible trademark protection. The medical background information delivered to users via the apps is often from universities, but it is unknown if royalties are paid to the universities and on which basis.

## 7. RESEARCH QUESTION, HYPOTHESES and METHODOLOGY

### 7.1. Research question

This research deals with the following research question, and hypothesis, triggered by the above data, survey and competitive analysis above:

***Research question: Can female fertility management mobile apps be sustainable and contribute to female health care*?**

### 7.2. Methodology

The above research question is interdisciplinary, and, as stated in Section 2, the mobile app suppliers have refused any direct interaction providing primary source data.

The chosen methodology is to formulate a few specific hypotheses characterizing the essential aspects of the research question above, and to validate (or not) each hypothesis by a combination of theoretical concepts, empirical data, and business analysis.

### 7.3. Researched hypotheses

The 7 hypotheses to be analyzed are the following:

a. Hypothesis 1: Female fertility management mobile apps with an educational dimension can assist in other female health care issues, such as prevention. YES/NO.
b. Hypothesis 2: The business generated by female fertility management apps can be extended to other female gynecological illnesses, and if YES, which exactly. YES/NO.
c. Hypothesis 3: Women user groups are reluctant to rely on female fertility management mobile apps functionality; if YES, which groups. YES/NO.
d. Hypothesis 4: Which are the exact measures of medical impact of female fertility management mobile apps? MEASURES.
e. Hypothesis 5: Benefit/cost to end user female from using a female fertility management mobile app is sufficient. YES/NO.

f. Hypothesis 6: A female fertility management platform is financially sustainable for a supplier. YES/NO.

g. Hypothesis 7: Female fertility management mobile apps, can, after adaptation, also help educate women about hormonal balance for pre-menstrual and after menopause phases. YES/NO.

## 8. HYPOTHESIS CHARACTERIZATION

### 8.1. Hypothesis 1: Female fertility management mobile apps with an educational dimension can assist in other female health care issues, such as prevention. YES/NO.

The female fertility management mobile app functionality analysis (see Tables 1 and 2) allow to identify two categories of functionality contributing to user education:

- Educational contribution to obstetrics and gynecology care;
- Educational contribution to other medical female health care issues such as prevention.

The mobile apps should eventually allow to educate users in these categories.

Obstetric diagnoses

- Anorexia and the menstrual cycle;
- Zika virus effects (Zika is widespread in Africa);
- Migraine headaches linked to the menstrual cycle;
- Genetic evidence for PMDD （premenstrual dysphoric disorder);
- Acute cystitis;
- Premenstrual magnification: Mental health conditions and PMS

- Spot and protection from sexually transmitted infections (STI) incl. HIV (see Figure 6 a geographical distribution of HIV, and in Figure 7 the number of resulting deaths).

Gynecology diagnoses

- Uterine polyps: a common reason for irregular bleeding around menopause;
- Menopause;
- Pregnancy symptoms.

Other diagnoses

Female fertility management mobile apps could assist in other educational female health care issues, such as prevention: especially against the Zika virus and sexual infections, as well as some types of migraine (to distinguish those due or not to fertility periods). Figures 6 and 7 about Africa give HIV and death statistics, highlighting the ultimate importance of allowing female health apps to combat such other diagnoses upfront.

*Conclusion: YES, if the mobile app has a feedback and diagnostic functionality. Female fertility management mobile apps which have functionalities in the two above diagnostic domains, by collecting data linked to the stated diagnostic classes, and providing feedback to female users, can provide significant educational value to their users. Because of the link to STI and HIV, and the significant likelihoods of these illnesses in Africa, the educational impact is especially high in Africa. Sex education including partners as well as parties corresponding to the woman's life stage are side effects.*

**Figure 6**: WHO 2016 regional distribution of adults and children living with HIV (WHO, 2017).

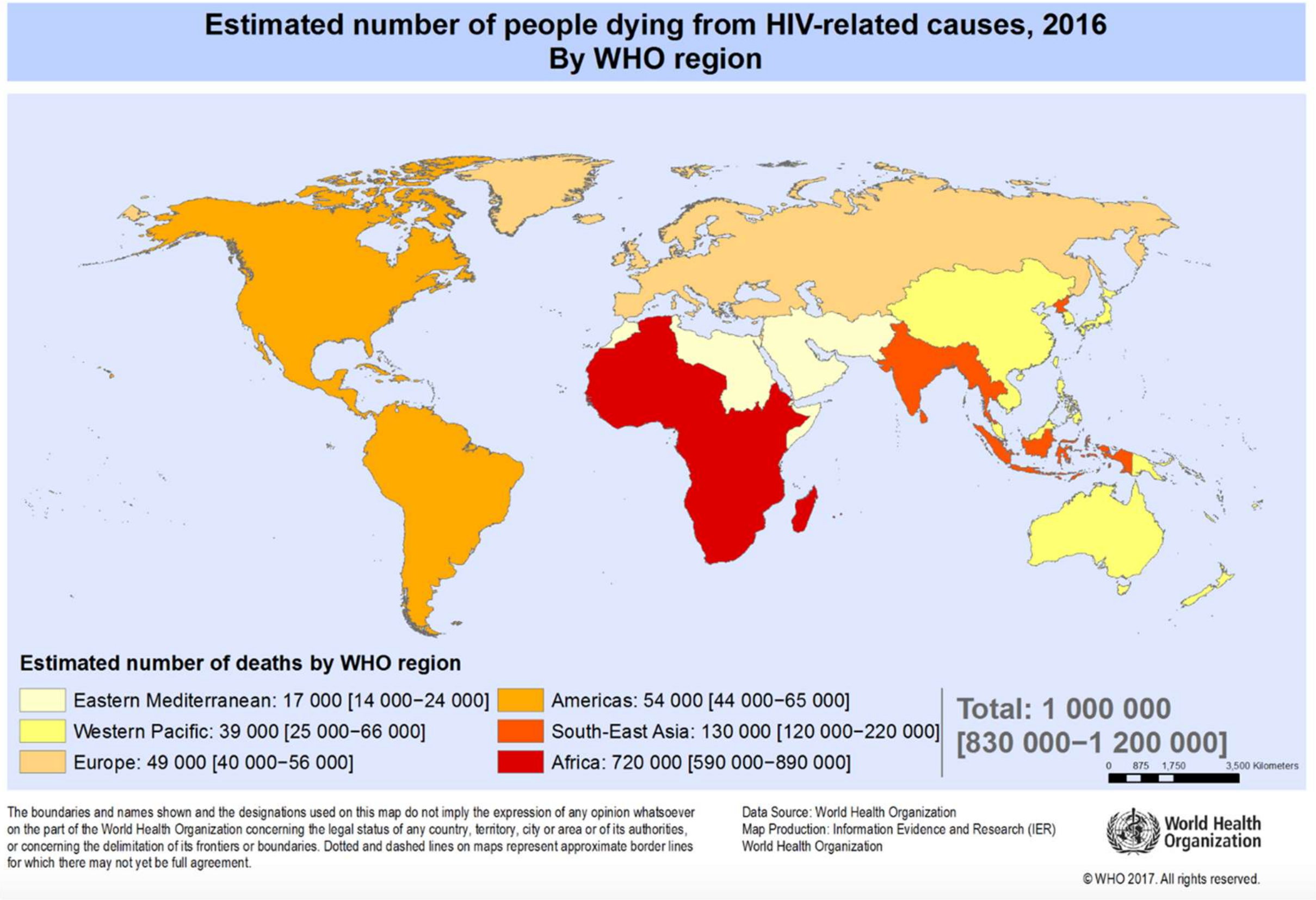

**Figure 7**: WHO 2016 data on leading causes of death in African region (WHO, 2018).

**Global Health Estimates 2016: Estimated deaths by cause and region, 2000 and 2016**

**Regional grouping:** WHO regions
**Region:** WHO Africa Region
**Year:** 2016

| Code | Cause of death | Both sexes | Male | Female |
|---|---|---|---|---|
| | *Sex* | Both sexes | Male | Female |
| | *Age group* | Total (all ages) | | |
| | Population (thousands) | 1,019,920 | 508,849 | 511,071 |
| 0 | **All Causes** | **8,845,104** | **4,639,627** | **4,205,477** |
| 10 | **I. Communicable, maternal, perinatal and nutritional conditions** | **4,951,786** | **2,574,269** | **2,377,517** |
| 20 | **A. Infectious and parasitic diseases** | **2,728,708** | **1,467,402** | **1,261,306** |
| 30 | 1. Tuberculosis | 405,496 | 253,537 | 151,960 |
| 40 | 2. STDs excluding HIV | 62,697 | 26,868 | 35,829 |
| 50 | a. Syphilis | 60,146 | 26,792 | 33,354 |
| 60 | b. Chlamydia | 508 | 0 | 508 |
| 70 | c. Gonorrhoea | 1,325 | 49 | 1,276 |
| 80 | d. Trichomoniasis | 0 | 0 | 0 |
| 85 | e. Genital herpes | 0 | 0 | 0 |
| 90 | f. Other STDs | 719 | 28 | 691 |
| 100 | 3. HIV/AIDS | 718,800 | 385,936 | 332,864 |
| 110 | 4. Diarrhoeal diseases | 652,791 | 342,026 | 310,765 |
| 120 | 5. Childhood-cluster diseases | 77,287 | 40,589 | 36,697 |
| 130 | a. Whooping cough | 6,516 | 2,641 | 3,875 |
| 140 | b. Diphtheria | 1,048 | 534 | 514 |
| 150 | c. Measles | 45,736 | 21,886 | 23,850 |
| 160 | d. Tetanus | 23,987 | 15,528 | 8,459 |
| 170 | 6. Meningitis | 186,069 | 100,113 | 85,956 |
| 180 | 7. Encephalitis | 9,482 | 5,280 | 4,202 |
| 185 | 8. Hepatitis | 31,340 | 15,753 | 15,587 |
| 186 | a. Acute hepatitis A | 733 | 431 | 302 |
| 190 | b. Acute hepatitis B | 24,085 | 12,034 | 12,051 |
| 200 | c. Acute hepatitis C | 743 | 355 | 389 |
| 205 | d. Acute hepatitis E | 5,779 | 2,933 | 2,845 |

## 8.2. Hypothesis 2: The business generated by female fertility management apps can be extended to other female gynecological illnesses, and if YES, which exactly. YES/NO.

This hypothesis deals with some extended medical benefits other than educational (Hypothesis 1). The female fertility management mobile apps, which also collect temperature data, can spot and track some other gynecological diagnoses than those considered above. This includes especially users reporting about symptoms of: uterine polyps, tumor, endometriosis, uterine fibroid. All these diagnoses might sometimes be an obstacle for female health, reducing fertility and causing sometimes pains.

If one or more of the symptoms the female user experiences before or during her period correspond to a menstrual cycle "disorder”, the mobile app allows to report them, thus flagging a risk for a specific diagnosis such as (Unnamed, 2018a):

- abnormal uterine bleeding (AUB), which may include heavy menstrual bleeding, no menstrual bleeding (amenorrhea) or bleeding between periods (irregular menstrual bleeding);
- dysmenorrhea (painful menstrual periods);
- premenstrual syndrome (PMS);
- premenstrual dysphonic disorder (PMDD).

*Conclusion: YES, if the mobile app has slightly enhanced data collection and a feedback diagnostic capability.*

*The indicated symptoms correspond to those indications usually collected by the female users, thus the mobile app can provide linked diagnostic capabilities beyond fertility management. This both increases the related business opportunity, but also the need to certify the mobile apps for these other female health diagnoses.*

### 8.3. Hypothesis 3: Women user groups are reluctant to rely on female fertility management mobile apps functionality; if YES, which groups. YES/NO.

From the literature survey references esp., Sections 5.2 and 5.3 have been identified very different and not always agreed upon bottlenecks creating a user reluctance:

-first the ultimate usefulness to the female user: avoiding getting pregnant, getting pregnant, or some pain management goal; thus, the data entry load and dependency on the mobile app (and time required) become a drain;

-next many women are afraid to let a third party handle their data and intimacy; the opinions on this are split;

-also, the user interface design may turn off some user, as experimental surveys have shown that some users want separate calendars for each month, while others want an overview to spot cycle variance;

-finally, the smartphone versions play a role, as some Android version users want to have similar input functions as IOS by finger tapping.

*Conclusion: UNDECIDED highly dependent on female end user personality.*

*The different identified causes of reluctance are very individually and culturally dependent. They are also very different, so those user groups showing reluctance may combine different reasons. This implies that user adoption of a female fertility mobile app may be very variable by location, culture, social level and education level of the user, as well as values. Also, the analysis shows that, a given female user may abandon or only use the mobile app temporarily.*

### 8.4. Hypothesis 4: Which are the exact measures of medical impact of female fertility management mobile apps? MEASURES.

Depending on the functionality of the female fertility management mobile app, there will be different sets of impact measures and quality criteria. The literature survey Section 5. shows that the lack of extensive testing and absence of medical certification can lead to inconsistent overall impact measure outcomes. Table 4 below attempts to set quality measures and impact measures for each of the functionalities identified in Section 3, Table 1. Because of the limited extent of the survey group described in Appendix 1, and the refusal of app suppliers to contribute data, it is impossible at this stage to quantify the proposed quality and impact metrics, although such should be a subject for further research.

**Table 4: Quality and impact measures from user point of view**

| ***Functionality*** | ***Description*** | ***Proposed Quality measure*** | ***Proposed Impact measure*** |
|---|---|---|---|
| ***1*** | *Menstruation prediction control, checking* | ***Variance of error on first day, and variance on error of last day*** | ***Two-week ahead prediction error on period start date, and duration; reduction in leave-of-absence days from work or home chores*** |
| ***2*** | *Ovulation prediction checking* | ***Variance of error on first day, and variance on error of last day*** | ***Two-week ahead prediction error on ovulation period start date, and duration;*** |
| ***3*** | *Basal body temperature* | ***Delay in predicting a 3 % deviation from normal body temperature*** | ***Reduction in leave-of-absence days from work or home chores*** |
| ***4*** | *Making note of specific physical condition* | ***Awareness or cognition for self-physical condition (might be controlled)*** | ***Reduction in leave-of-absence days from work or home chores*** |
| ***5*** | *Diet functionality such as weight control* | ***Delay in predicting a 5 % deviation from normal weight (or combined height-weight index)*** | ***Providing basic cal/protein/lipid levels with 10 % tolerated inaccuracy for nutritional planning*** |
| ***6*** | *Predict Notification for next menstruation* | ***Zero error on notification and zero error on possible*** | ***Subject to end user authorization,*** |
| | | ***notification retransmission*** | ***error-free reporting of menstruation date prediction to medical profession*** |
| ***7*** | *Cycle extent education (information about sleepiness, irritation, etc..)* | ***Awareness or cognition for self-physical condition (might be controlled)*** | ***Reduction in leave-of-absence days from work or home chores*** |
| ***8*** | *Health log (food, feelings, sex drive status)* | ***Possibilities for comparing each month´s statement*** | ***Adopt one of approved overall health assessment scales, like the Stanford scale (Bruce B., Fries J.F., 2003).*** |
| ***9*** | *Symptoms (headache, digestion)* | ***Awareness or cognition for self-physical condition (might be controlled)*** | ***Reduction in leave-of-absence days from work or home chores*** |
| ***10*** | *Sexual activity* | ***Rate of pregnancies occurring because of errors in fertility or ovulation dates, assuming no contraceptives*** | ***Rate of unwanted pregnancies*** |
| ***11*** | *Medication log* | ***Measurable effects to menstruation conditions*** | ***Reduction in leave-of-absence days from work or home chores*** |
| ***12*** | *Fertility assistance* | ***Pre-Infertility treatment and reducing medical supports*** | ***Reduction in leave-of-absence days from work or home chores*** |
| ***13*** | *Information, Q&A, articles* | ***Awareness and knowledge level by users*** | ***Reduction in hospital consultations for general menstruation information*** |
| ***14*** | *Messaging and chat in app* | ***Reduction of psychological anxiety and information sharing*** | ***Standard social network interaction levels such as number*** |
| | | | ***of moderated responses to a given topic*** |
| ***15*** | *Pregnant women physical training program* | ***Enhance physical strength and endurance of pregnant women*** | ***Reduction in leave-of-absence days from work or home chores*** |
| ***16*** | *Emotional mood* | ***Reduction in anxiety level*** | ***Reduction in leave-of-absence days from work or home chores*** |
| ***17*** | *Pill diaries* | ***Measurable effects on menstruation conditions esp. duration of pains*** | ***Reduction in leave-of-absence days from work or home chores*** |
| ***18*** | *Vaginal condition (vaginal secretions)* | ***Awareness and protective measures against sexually transmitted infections*** | ***Reduction in hospital treatments for sexually transmitted diseases*** |

The female fertility planning app is not isolated in the public health system: not only it gives basic information e.g. about menstruation, but the app collected data provides possibly inputs to so-called primary care function.

In this way, the female fertility management mobile apps give society a great medical impact as part of preventive or primary care or helping manage work presence and thus leave of absence claims.

Nevertheless, the quality and impact can only be assessed if data are provided by suppliers for validation and certification.

In addition, medical reform, planning development, and policies towards appropriate social conditions within each country at government level are desirable. In the current era where private enterprises too are aiming for medical improvement at society level, they may take other avenues than the public health system.

Although smartphones are very widely distributed in developing countries, the awareness, now nonexistent there, of female fertility management mobile apps, would help women 's health but also social standing, proving the points explained in Figure 3 (gender and women's empowerment)

*Conclusion: MEASURES (quality and impact) PROVIDED but values not yet available.*

*Table 4 provides quality and impact measures which could be collected if a number of conditions are met. First, they apply differently as not all female fertility management platforms have the same functionality, so calibration and verification are mandatory. Next, by certification and incentives subject to data privacy constraints, the data should be anonymized to allow measures to be assessed. The app service suppliers should abandon their sole ownership, and resale business model. Finally, suppliers should compete on such measures, a stage they are not yet prepared for.*

### 8.5. Hypothesis 5: Benefit/cost to end user female from using a female fertility management mobile app is sufficient. YES/NO.

This hypothesis deals only with the female end user point of view. There are almost no market analyses of female fertility management mobile apps, but the French Cosmopolitan magazine (Roche J., 2018) as done an extensive analysis which contains precious data, probably more trustworthy than the very few information items provided by the suppliers of Table 1.

Benefits are obviously linked to the number of users. However, as the benefits of female fertility management obviously require tracking over a significant period, all information by suppliers or Google Analytics on one-time downloads or one-time users does not demonstrate benefits to end users.

Therefore, for those mobile apps which offer regular subscriptions (paying or not), the number of paying active subscribers is a far better indication, and French Cosmopolitan (Roche J.,2018) gives the following data:

1) Clue has only 46 800 subscribers as of now in 2018;

2) "Cycles menstruels-suivi", is equivalent to Clue, but simpler to use; the number of users is not indicated;

3) "Flo Period tracker", with graphs and different functionality; it has 88 300 subscribers as of now in 2018; fancy user interface;

4) "Eve" offers not only menstrual cycles info, plus many other sexual information functions for women; it was named in 2016 " most innovative app". It also has a community chat functionality; unknown number of users;

5) "Glow" released a new “Ruby” application much more focused on sex education that helps young generation women’s pregnancy, correcting sexual knowledge.

The literature survey, and especially Section 5.2. with the use of the Mobile Application Rating Scale (MARS) shows that end user’s perceived benefits from usage of health apps includes greater self-awareness of one’s condition, easier integration of self-management in daily life, ability to send data to allied health professionals without repeated visits, the ability to view historical data without visiting a doctor, social motivation to improve fitness, and desire to customize app features to suit individual needs. Participants also expressed greater control of their condition, in this case, menstrual problems.

To further analyze benefits and costs, one author and a very limited survey group of 3 users has carried out an in-situ experimental evaluation of benefits and costs from a female end user point of view. The results are provided in Appendix 1. This experimental research shows that benefits are perceived, but not unanimously accepted, and that sometimes the expected benefits are more around female health care in general, than around fertility period planning. Thus, the demand picture from Appendix 1 indicates a wish for a broader scope than the supply, to include such functionality as pre-menstrual syndrome, breast cancer detection, and anti-ageing.

Regarding costs to female end users, they can be analyzed in terms of monetary costs, of time spent, and of perceived “costs” or uneasiness in sharing data with others. Depending upon the mobile apps from Table 1, and excluding mobile device cost and mobile subscription costs, the female fertility mobile app subscriber costs range from 0,5 to 10 Euros/month, in that higher rates are for now not accepted by end users. Appendix 1 experimental research confirms that essentially free use is desired, and that the few convinced users would at most pay 0,5 Euros/ month. Some European platforms offer one-time usage “pay-by-use” fees in the 0,50-1 Euro range, but they seem to have a very limited following. It is only in the United States that premium subscriptions like Flo Period Tracker fetch almost 10 USD/month due probably to high levels of Internet app addiction. In Japan, Luna Luna attracts affluent women with monthly subscription fees of either 194 Yen or 400 Yen (3, 17 Euros)/month, riding on a strong wish to have a baby in a country with low fertility.

The data entry time is estimated from the experimental research in Appendix 1 at 6 minutes/day for 6 days, that is 36 min/month, and mobile app consultation time depends on end user personality and levels of anxiety about menstruation, with 10 min/day browsing during 7 days, i.e. in total 70 min/month, depending on the ease of use of the user interface.

It is impossible to assess the uneasiness in sharing data with third parties, and Section 5.2 shows in another study that there are shared views on this (see Section 6.2.).

*Conclusion: YES, for few regular users, NO for other users.*

*The analysis above shows that for regular users there are clear benefits and that the perceived costs are negligible. In case regular users must pay a subscription fee or a one-time fee, at the*

*current levels practiced, the benefits outweigh these monetary costs. By far the preferred user mode is free use of the app.*

*At the same time, there is a huge discrepancy between announced user bases by suppliers (for Clue, 2.5 M claimed users, with only 46 800 verified paying users) tending to indicate that once awareness is created, there is a significant loss over time and non-paying users dominate.*

### 8.6. Hypothesis 6: A female fertility management platform is financially sustainable for a supplier. YES/NO.

This hypothesis deals only with the financial sustainability of the supply of a female fertility management service platform. In this analysis we reuse valuation estimates from the financial information in Section 6.5.

The largest company now "Flo Period tracker", according to (Roche J.,2018), is 10 years old and has about 80 000 registered paying users. The overall revenue basis is the analysis below is based on a mix of subscriptions (average service revenue per one-month user session of 0,25 Euros, determined in Section 6.5.) and add /product / data sales revenue to the supplier, assuming the end user's fertility information updates session period to last 6 months (e.g. in relation to pregnancy or possible pregnancy). This produces a monthly service revenue of:

ServiceRev=0,25 Euros x N Euros/month, where N is number of paying registered users

As the largest company ("Flo Period Tracker") has about 80 000 registered users the largest service revenue in the market covered by the suppliers of Table 1 is:

Max (ServiceRev) = 0,25 Euros x 80 000 subscribers = 20 000 Euros/month

Assuming now, for a mature supplier in that segment to have about equivalent monthly revenues from advertisements carried by the app, from data resale (if allowed), from products or courses, etc.., the largest total revenue in the market covered by the suppliers of Table 1 is:

Max (Rev) = 2 x Max (ServiceRev) = 40 000 Euros/Month.

It is reasonable to assume, for that female fertility class of mobile app services, to assume fixed costs as follows:

a. IT system operations or cloud capacity lease costs for a mobile information service (this is higher than serving fixed Internet users): 10 000 Euros/month (usually charged by mobile operators or a virtual mobile operators); these capacity costs must support the usage by the very many nonpaying and sporadic users!

b. Marketing costs: 5000 Euros/month

c. General, software updates and administration accounting etc.: 5 000 Euros/month

It is noted from Table 2 that some suppliers have many multilingual versions, which add software internationalization costs to the base app development costs.

This leaves: Max (Rev) - 10 000 – 5000-5000= 0,5 x N – 20 000 Euros /month for salaries, personnel charges and profits.

The breakeven number N of paying subscribers is thus 40 000 users (solution of 0,5xN-20 000=0), assuming no salaries and no profits! “Flo Period Tracker” with 88 000 subscribers may be profitable if its running personnel and finance costs are very moderate. It charges 9,9 USD/month for its premium service or 49,9 USD/year. If all “Flo Period tracker” registered subscribers pay the annual premium service, one can determine that its staff size is no more than 20 permanent employees on a yearly basis.

But with its 46 800 registered paying users in 2017 the Clue app, BioWink (Section 2) is probably loss making! To be more precise, BioWink collected significant seed capital since 2013; the founder claims a staff size of 50 people (Bloomberg, 2018), which correspond to a cash burn rate of about 7,5 Million Euros/year (assuming man x year cost of 150 000 Euros/year with charges). Although collected seed capital declarations vary between 10 MEuros and 30 M$, and although ramp up/down of staff is unknown (see Section 2), it is highly likely that BioWink not only is loss making but is also about to run out of capital in 2017-2018. Also, the analysis of Section 6.5 indicates a likely negative return on investment for investors.

*Conclusion: NO*

*A female fertility management service platform is very likely not to be financially viable for a supplier, and thus exhibits very high risks for an investor (see financial valuation analysis in Section 6.5). This business analysis is of course subject to all reservations, as none of the female fertility management suppliers of Table 1 provided real data and anyway data would not be public. That is why it is mostly an informed and educated guess using closely related information. It is also impossible to assess the upside to business from selling data, from advertising income, and from short duration users.*

*Nevertheless, the business analysis is in sharp contrast with an estimated valuation by investors of 213 Euros of seed investment value per user in a stable phase (see Section 6.5).*

It is therefore for a fertility APP company necessary to consider following options:

1. Charge a subscription fee /month, for so-called "premium services", but what is then the fall in number of users? Certain Clue features will be part of a "pro version" or "Clue membership", where you pay a monthly fee to receive access to new features while also supporting the BioWink company and vision;
2. On the same platform, host more services than just fertility planning, to increase the values of N for different services, and amortize fixed costs;
3. Offer custom personalized services, for an individual hourly-based fee, e.g. on gynecology, sex, as some competitors of Clue do;
4. Sell itself after a startup phase, to a larger eHealth company; return to founders can be estimated at 5 (Euros valuation per subscriber) x N (see Section 6.5) which is still bad business to the founders as Max (N) is 88 000.

## 8.7. Hypothesis 7: Female fertility management mobile apps, can, after adaptation, also help educate women about hormonal balance for pre-menstrual and after menopause phases. YES/NO.

Female health care is necessary before and after menstruation (sexual maturity), that is, women before menarche (puberty) and women before and after menopause. For women before menarche and early menarche, symptoms peculiar to the menarche (irregular menstruation or rough skin, disorder of appetite, pimples, etc.) can be cited. It is a prudence period and a sense of pleasure and family support is desirable. However, differences in education may arise for various reasons. The ideal is proper support by the family, but sometimes there are misunderstandings from cultural and environmental reasons. It is also so that sometimes in some cultures, one of the purposes is to share the proper correct knowledge with family members other than the users. Of course, for women before and after menopause too, there may occur unique symptoms (irregular menstruation by menopause, or disorder of hormonal balance, accompanying physical symptoms and work absences: menopausal disorder). A female fertility management mobile app functionality (Tables 1 and 2) may be extended to include, from the reported indications, such disorders.

Because the same end user date would be required, with a few additional ones, the above observation opens the opportunity to address more broadly female healthcare, aiming at establishing knowledge of about each female age group, measuring information differences between age groups, and by extending functionality first for premenstrual women, then pre and postmenopausal women.

*Conclusion: YES, if mobile app offers a feedback and advisory functionality.*

*Although this is only a potential extension to the functionalities of female fertility management mobile apps, addressing premenstrual then menopausal women, and their fitness for work, appears to be as an educational tool, quite doable.*

## 9. CONCLUSION AND ANSWER TO RESEARCH QUESTION

Have been established in Section 8, with limitations owing to information access from suppliers (see Section 2), the following outcomes of the hypothesis related to the research question (summarized here in Table 5):

**Table 5: Summary of hypotheses testing outcomes with conditions**

| HYPOTHESIS | Hypothesis testing outcome | CONDITION |
|---|---|---|
| 1 | YES | *If the mobile app has a feedback and diagnostic functionality* |
| 2 | YES | *If the mobile app has slightly enhanced data collection and a feedback diagnostic capability* |
| 3 | UNDECIDED | *Highly dependent on female end user personality* |
| 4 | MEASURES PROVIDED | *But values not yet available* |
| 5 | *YES, for few regular users, NO for other users* | *The conditions are very personal, and clarified by the experimental research in Appendix 1* |
| 6 | NO | *Financial sustainability measures include: cash flow which must be positive, return on investment which must be higher than other mobile services, and founder's goals* |
| 7 | YES | *If the mobile app offers a feedback and advisory functionality* |

To these outcomes, we apply multiple-criteria Pareto voting, with no weighing to determine the dominating aggregate answer from the seven sub-answers (Arnold, 2012).

To the research question: **"Can female fertility management mobile apps be sustainable and contribute to female health?", the answer Is a <u>moderate yes</u>, provided the following conditions:**

-the female fertility management mobile app must offer feedback and advisory/diagnostic capabilities, even if limited, to be able to offer significant benefits to end users in cl. Education, labor medicine and public health;

-such mobile apps are not business-wise sustainable, under their present functionality and organizational set-up;

-user adoption is highly dependent on female end user personality, and its development is slowed down by lack of transparent certified quality and impact data from providers;

-areas for smooth functionality enhancement towards wider female eHealth have been identified.

## 10. FURTHER BUSINESS OPPORTUNITIES AND INNOVATION POTENTIAL

The academic literature and industry discussion about the evolution of female fertility management apps, is either scarce or superficial (Section 5 and 6 resp.). This is largely due to the reluctance of suppliers to allow to analyze real impacts and specific personalized user queries.
In addition, from the above analysis emerge the following further research topics, innovative ideas for business consolidation and expansion:

### 10.1. Future research

-Contributions of female fertility management apps in developing countries (e.g. (National Health Planning of India, 2017);

-How mass data is used and analyzed for scientific research e.g. in gynecology;

-What could be the next business model for female fertility management app suppliers?

-How female users would organize themselves, between themselves, to provide the best feedback of their experiences?

-Clue has called on Apple to develop a device to put inside the body on a molecular level (Shead, 2017), but this story could be just a pitch to get marketing attention wrongly triggered by Apple Watch announcements.

### 10.2. Innovative ideas and functionality expansion

- Education for all age generations of women, e.g. to post-menstruation periods as well to post-birth women especially if they are in the workforce;
- Enhanced collaborations and information sharing with other organizations, such as schools, hospitals, pharmaceutical companies, body training/fitness companies, advertisement companies.

Although not revealed by the experimental research in Appendix 1, which catered to women only, a realistic opportunity exists *for male fertility management!* As female fertility obeys to other physiological processes, the overlap with the apps in Table 1 is minimal. However, the same supplier could try to cater to male and females under the same marketing and branding concept. This combination is not yet available, although male fertility risks are the specialty of a very interesting French start-up Spartan (Spartan, 2016). It deals with the electromagnetic effects on male organs, by offering not only textile technology but also a mobile app measuring the induced electromagnetic field (without and with protecting textiles) in view of the surrounding electromagnetic field due mostly to WiFi irradiation in different frequency bands, and to a lesser extent 3G/4G exposure, due to male user's huge dependence on smartphones. The protective underwear has sold as of end of 2017 over 10 000 units at the unit cost of 42 Euros.

### 10.3. Marketing ideas for business consolidation & expansion

As revealed by the Hypothesis 6 analysis, the female fertility management mobile app suppliers must engage and partner with other sectors, and not stay mostly independent. This applies to knowledge validation, market outreach, but also for branding and advertisement revenue. The prime sectors to target are:

1. Pharmaceutical companies (for contraceptive and external medicine);
2. Hospitals (to help female patients and personnel in their ovulation cycle control, and to strengthen protection from venereal diseases);
3. Educational organizations (educational information to young students);
4. Training and fitness organizations (to propose exercise for pregnant women in collaboration with medical institutions);
5. Media advertising companies to create awareness if not diversification revenue (see Section 6.5. for examples).

## APPENDIX 1: A LIMITED EXPERIMENTAL COST/ BENEFIT ANALYSIS FROM FEMALE END USER PERSPECTIVE

Over the period March 2018 to July 2018, one author assembled a small group of 3 women (User 1: age 33, User A: age 23, User B: age 40) to specify the demand side of female fertility management, and to assess some benefits as well as extensions. All members provided data and rated one female fertility management app (Clue) (see results in Table 6).

The frequency of debriefing meetings was once per month (one Face-to-face, the other via Skype). All survey members were using the same mobile app, Clue, in different languages (English, and Japanese); the functionality in Clue was the same.

Questions raised were concerning app design, registration of personal data, usability of the app, customization functionality in the app.

Unsolicited comments received were about: availability of educational information on anti-ageing practices, difficulty in filling in time the required user data, inadequate or odd language translations in apps, availability of education about pre-menstrual period (PSM), procedure to check breast in view of breast cancer risk.

There were few comments on the period prediction errors, ranging from 3 days (50 % probability), to 5 days (50 % probability), showing that the Cue app cannot predict more accurately, in view of body condition and activities.

User data were entered over several days duration, during the menstruation period only (typically 7 days); data entry time spent ranges from 3 min to 10 min, with an average of 6 min.

Users were looking daily at the Clue app, starting typically from one week before the expected menstruation; browsing of the information typically takes 10 min. The output of the app includes bar charts. No user was printing any results not even the calendar.

In this user group, no users used the paying version of the app, only the free edition (User 1 and User A on Android, User B on iPhone). Access only by WiFi from home, typically at night before sleep.

Main benefit mentioned was to bring a knowledge and educational dimension to a natural process. The other was to help to resupply sanitary devices like tampons, for those users who could not remember to do. In case of irregular pain, the app gives some explanations, and likewise if stronger pain level occurs due to physical stress, or environmental stress. Clue server was never down.

The user group was not happy with the app design: the calendar is difficult to read, and the months are all adjacent. User compared Clue with Luna Luna app, which has a more attractive layout for women, while Clue is too cold. For some users, the pop-up notifications did not arrive at all, and at times the app would not open.

The younger members in the user group trust the app and intend to continue to use the app if it is free. If the monthly subscription would reach 0,5 Euros/month, half of the continuing users would stop, and the remaining half would stop at 1,5 Euros/month. The older member of the survey group switched to another free app. If a user in the group would get pregnant, initially she will check both the app and a doctor to check on delay, then rely on doctor primarily.

If functionality is extended beyond female fertility management, the preferred additional functionalities should deal with PMS, breast cancer prevention, and anti-ageing.

**Table 6: Experimental research results on a female fertility planning tool, from user´s point of view.** A scaled assessment was made, with the following scale values:

0: Not needed or useless, 1: Insufficient, 2: Sufficient, 3: Good

| **Experimental data from end user** | User 1 | User A | User B |
|---|---|---|---|
| **Assessment of benefits of the Clue app** | | | |
| 1 registration of personal data by user | 2 | 3 | 2 |
| 2 registration of menstruation data by user | 3 | 3 | 2 |
| 3 usage loyalty over time | 3 | 2 | 2 |
| 4 value of anti-aging information | 2 | 1 | 2 |
| 5 notification messaging about due date for taking pill | 0 | 3 | 2 |
| 6 suggested date for checking breasts | 0 | 2 | 0 |
| 7 value of pre-menstrual syndrome information | 3 | 3 | 3 |
| 8 customizing functionalities | 3 | 3 | 3 |
| **Costs** | | | |
| Cost paid/month | Free version | Free version | Free version |
| Assessment of usage cost if 0,50 Euros/month | 1 | 1 | 0 |
| **Personal data entry time** | | | |
| Assessment of date entry time used | 2 | 2 | 2 |
| Data entry time (min) ×days in one month | 3min x 6 days | 5min x 7 days | 10min×7days |
| **Total scores** | 19 | 23 | 18 |
| **Total average score (excluding 0)** | **2,4** | **2,3** | **2,3** |

The overall ratings by all users are consistent, in the range 2,3 to 2,4, which means that the Clue app was sufficient, but not good, at providing to users the functionality demanded.